\documentstyle[12pt]{article}
\begin{document}

\begin{center}
{\bf \Large Envelope Solitons of Nonlinear Schr\"{o}dinger
Equation with an Anti-cubic Nonlinearity\footnote{Work
supported by CRUI (Conferenza dei Rettori delle
Universit\'{a} Italiane) and DAAD (Deutscher Akademischer
Austauschdienst) within the research programm "Vigoni"
between University of Napoli Federico II and University
of Bayreuth  (Project {\it Nonlinear Phenomena,
Instability and Macroscopic Coherence in Charged-Particle
Beam Dynamics})}}
\end{center}

\medskip

\begin{center}{\bf \Large R. Fedele,
H. Schamel\footnote{Physikalisches
Institut, Universit\"{a}t Bayreuth, D-95440 Bayreuth,
Germany.\\ Email: hans.schamel@uni-bayreuth.de}, V.I.
Karpman\footnote{Racah Institute of Physics, Hebrew
University, Jerusalem 91904, Israel.\\ Email:
karpman@vms.huji.ac.il}, P.K. Shukla\footnote{Institut
f\"{u}r Theoretische Physik IV, Fakult\"{a}t f\"{u}r Physik und
Astronomie, Ruhr-Universit\"{a}t Bochum, D-44780 Bochum,
Germany and Department of Plasma Physics, Ume{\aa}
~University, SE-90187 Ume{\aa}, Sweden.\\ Email:
ps@tp4.ruhr-uni-bochum.de}}
\end{center}
\begin{center}{\it Dipartimento di Scienze Fisiche, Universit\`{a}
Federico II and INFN, \\ Complesso Universitario di M.S. Angelo,\\ Via
Cintia, I-80126 Napoli, Italy\\ Email: renato.fedele@na.infn.it}
\end{center}

\bigskip

\begin{center}
{\bf Abstract}
\end{center}

{On the basis of a recently-proposed method to find
solitary solutions of generalized nonlinear Schr\"{o}dinger
equations \cite{1}-\cite{3}, the existence of an envelope
solitonlike solutions of a nonlinear Schr\"{o}dinger equation
containing an anti-cubic nonlinearity ($|\Psi|^{-4}\Psi
$) plus a "regular" nonlinear part is investigated. In
particular, in case the regular nonlinear part consists
of a sum of a cubic and a quintic nonlinearities (i.e.
$q_1|\Psi|^2\Psi+q_2|\Psi|^4\Psi$), an upper-shifted
bright envelope solitonlike solution is explicitly
found.}

\bigskip\bigskip
\noindent
{\small PACS number(s): 52.35.Mw - Nonlinear waves;
05.45.Yv - Solitons;\\ 42.65 - Nonlinear optics; 67.57.Jj
- Collective modes in quantum fluids}

\medskip\medskip


\newpage
Let us consider the following generalized Korteweg-de
Vries equation (GKdVE)
\begin{equation}
a~{\frac{\partial u}{\partial s }}~-~G\left[u\right]~
{\frac{\partial u}{\partial
x}}~+~\frac{\nu^2}{4}~{\frac{\partial^3 u}{\partial
x^3}}~=~0~~~,
\label{gkdve}
\end{equation}
where $a$ and $\nu$ are real constants, and
$G\left[u\right]$ is a real functional of $u$, and the
following generalized nonlinear Schr\"{o}dinger equation
(GNLSE)
\begin{equation}
i\alpha{\frac{\partial\Psi}{\partial
s}}~+~{\frac{\alpha^2}{2}}{\frac{
\partial^2\Psi}{\partial x^2}}~-
~U\left[|\Psi|^2\right]\Psi~=~0~~~,
\label{gnlse}
\end{equation}
where $U\left[|\Psi|^2\right]$ is a real functional of
$|\Psi|^2$ and $\alpha$ is a real constant. \newline
Recently, it has been shown that a correspondence between
(\ref{gkdve}) and (\ref{gnlse}) has been constructed in
such a way that, provided that the following equations
are satisfied, i.e.
\begin{equation}
\nu~=~\alpha
\label{nu-alpha}
\end{equation}
\begin{equation}
u_0 a~=~-u_{0}^{2}~+~2c_0
\label{}
\end{equation}
\begin{equation}
G\left[u\right]~=~u\frac{dU\left[u\right]}{du}~
+~2U\left[u\right]~~~,
\label{functional-G}
\end{equation}
if $u(x-u_0 s)\equiv u(\xi)$ ($u_0$ being a real
constant) is a non--negative stationary--profile solution
of (\ref{gkdve}) thus
\begin{equation}
\Psi (x,s)=\sqrt{u(\xi)}\exp\left\{\frac{i}{\alpha}
\left[\phi_0-\left(c_0 + u_{0}^{2}\right)s+u_0 x
+A_0\int\frac{d\xi}{u (\xi)}\right]\right\}~~~,
\label{Psi-solution}
\end{equation}
is a stationary--profile envelope solution of
(\ref{gnlse}), where $\phi_0$, $c_0$, and $A_0$ are real
constants (not all independent) \cite{2}. Note that
$u(\xi) = |\Psi (x,s)|^2$. This result has been
successfully applied to find new analytical solitary-wave
envelope solutions of some modified nonlinear Schr\"{o}dinger
equations (MNLSE) containing high--order nonlinearities
(with respect to the standard cubic nonlinearity)
\cite{2,3}.  In particular, analytical solutions in the
form of bright and dark/grey envelope solitonlike
solutions have been found for a MNLSE with cubic plus
quintic nonlinear terms, i.e.
\begin{equation}
{\cal U}\left(|\Psi|^2\right)
= q_1|\Psi|^2 + q_2|\Psi|^4~~~.
\label{cal-U}
\end{equation}
In fact, it has been shown that the following MNLSE
\begin{equation}
i\alpha{\frac{\partial\Psi}{\partial
s}}~+~{\frac{\alpha^2}{2}}{\frac{
\partial^2\Psi}{\partial x^2}}~-
~\left[q_1~|\Psi|^2~+~q_2~|\Psi|^4\right]\Psi~=~0~~~,
\label{mnlse-quartic-bis}
\end{equation}
has the following envelope solitonlike solutions
\cite{2}:
\begin{eqnarray}
\Psi (x,s)& = &\sqrt{\overline{u}~\left[1~+~\epsilon~\mbox{sech}
\left(\xi /\Delta\right)\right]}\exp\left\{\frac{i}{\alpha}
\left[\phi_0 - A s + u_0 x \right]\right\}{\times}\nonumber\\
& {\times} & \exp\left\{\frac{i B}{\alpha}
\left[\frac{\xi}{\Delta}+\frac{2\epsilon}{\sqrt{1-\epsilon^2}}
\mbox{arctan}\left(\frac{\left(\epsilon -1\right)\mbox{tanh}
\left(\xi /2\Delta\right)}{\sqrt{1-\epsilon^2}}\right)
\right]\right\}~~~,
\label{envelope-soliton-quartic}
\end{eqnarray}
where $\phi_0$ still plays the role of arbitrary
constant,
$$\epsilon ={\pm}\sqrt{1- 32|q_2|\left(u_0 -
V_0\right)^2/\left(3q_{1}^{2}\right)}~~~,$$
$$\Delta
=|\alpha |/\left(2\sqrt{2|E'_{0}|}\right)~~~,$$

$$E'_0 =
-3q_{1}^{2}/\left(64|q_2|\right)~+~\left(u_0 -V_0\right)^2/2~~~,$$
provided that $E'_0 <0$ and $q_2 <0$ and
$$
-\sqrt{\frac{3q_{1}^{2}}{32|q_2|}}+V_0 <u_0 <
\sqrt{\frac{3q_{1}^{2}}{32|q_2|}}+V_0~~~,
$$
\begin{equation}
A~=~\frac{15 q_{1}^{2}}{64|q_2|}+\frac{\left(u_0 -
V_0\right)^2}{2} +\frac{u_{0}^{2}}{2}~~~,
\label{A-def}
\end{equation}
and
\begin{equation}
B~=-\frac{|\alpha|\left(u_0 -
V_0\right)}{2\sqrt{2|E'_0|}}~~~~.
\label{B-def}
\end{equation}

The following four cases have been discussed \cite{2}

\medskip
\noindent
(a). $0<\epsilon <1$ ($u_0 -V_0\neq 0$):
$$u(\xi=0)=(1+\epsilon)\overline{u}~,~~~\mbox{and}~~~
\lim_{\xi\rightarrow {\pm}\infty}u(\xi)=\overline{u}$$
which corresponds to a bright soliton of maximum
amplitude $(1+\epsilon)\overline{u}$ and up-shifted by
the quantity $\overline{u}$ ( {\it up-shifted bright
soliton}).

\bigskip
\noindent
(b). $-1<\epsilon <0$ ($u_0 -V_0\neq 0$):
$$u(\xi=0)=(1-\epsilon)\overline{u}~,~~~\mbox{and}~~~
\lim_{\xi\rightarrow {\pm}\infty}u(\xi)=\overline{u}$$
which is a dark soliton with minimum amplitude
$(1-\epsilon)\overline{u}$ and reaching asimptotically
the upper limit $\overline{u}$ (standard {\it gray
soliton}).

\bigskip
\noindent
(c). $\epsilon =1$ ($u_0 -V_0 = 0$):
$$u(\xi=0)=2\overline{u}~,~~~\mbox{and}~~~
\lim_{\xi\rightarrow {\pm}\infty}u(\xi)=\overline{u}$$
which corresponds to a bright soliton of maximum
amplitude $2\overline{u}$ and up-shifted by the maximum
quantity $\overline{u}$ ({\it upper-shifted bright
soliton}).

\bigskip
\noindent
(d). $\epsilon =-1$ ($u_0 -V_0 = 0$):
$$u(\xi=0)= 0~,~~~\mbox{and}~~~
\lim_{\xi\rightarrow {\pm}\infty}u(\xi)=\overline{u}$$
which is a dark soliton (zero minimum amplitude),
reaching asimptotically the upper limit $\overline{u}$
(standard {\it dark soliton}).

In this paper, we use the method mentioned above to find
envelope solitonlike solutions of the following MNLSE
containing, besides the cubic and quintic nolinearities,
an "anti-cubic" nonlinearity (i.e. $|\Psi|^{-4}\Psi$),
namely
\begin{equation}
i\alpha{\frac{\partial\Psi}{\partial
s}}~+~{\frac{\alpha^2}{2}}{\frac{
\partial^2\Psi}{\partial x^2}}~-
~\left[Q_0~|\Psi|^{-4}~+~q_1~|\Psi|^2~+~q_2~|\Psi|^4\right]\Psi~=~0~~~,
\label{mnlse-anti-cubic}
\end{equation}
where $Q_0$ is a real constant. \newline It is easily
seen that Eq. (\ref{functional-G}) has the following
general solution
\begin{equation}
U\left[u\right]~=~\frac{1}{u^2}\left[K_0~+~
\int~G\left[u\right]~u~du\right]\,,
\label{functional-solution}
\end{equation}
where $K_0$ is an arbitrary real constant. However, once
the function $u(\xi)$ is a stationary profile of
(\ref{gkdve}) with
\begin{equation}
G\left[u\right]~=~3q_1~u~+~4q_2~u^2\,,
\label{functional-G-1}
\end{equation}
it follows that, provided that $Q_0 = K_0$,
\begin{equation}
{\cal U}\left[u\right]~= ~{1\over u^2
}\int~G\left[u\right]~u~du\,,
\label{functional-solution-1}
\end{equation}
where ${\cal U}$ is defined by (\ref{cal-U}).
Consequently, it seems that the mapping
(\ref{Psi-solution}) allows us to construct solutions of
the (\ref{mnlse-anti-cubic}) which have the functional
form of the ones satisfying the
(\ref{mnlse-quartic-bis}). The only restriction for
(\ref{mnlse-anti-cubic}) is that $u(\xi)$ must not vanish
somewhere, which implies that we have to impose that
${u(\xi)}$ be a positive solitonlike solution of the
MKdVE of the type (\ref{gkdve}) with the nonlinearity
given by (\ref{functional-G-1}). In fact, the vanishing
of $u$ corresponds to a divergence of the nonlinear
potential term $|\Psi|^{-4}$ in (\ref{mnlse-anti-cubic}).
This circumstance excludes the standard dark solitonlike
solutions.
\newline On the other hand, it can be shown that \cite{2}
\begin{equation}
-\frac{\nu^2}{2}\frac{d^2 u^{1/2}}{d\xi^2
}~+~\frac{Q_0}{u^{3/2}}~+~\frac{1}{u^{3/2}}~
\int~G\left[u\right]~u~du~=~\left(c_0~+~
\frac{u_{0}^{2}}{2}\right)u^{1/2}~-~\frac{A_{0}^{2}}{2u^{3/2}}~~~.
\label{u-equation}
\end{equation}
It is then clear from (\ref{u-equation}) that a family of
solitary wave solutions of (\ref{mnlse-anti-cubic}) can
be obtained by imposing the following condition
\begin{equation}
A_0~=~{\pm}\sqrt{-2Q_0}\,,
\label{A-zero-condition}
\end{equation}
which implies that such a kind of family of solution
exists for negative values of $Q_0$. Consequently, Eq.
(\ref{u-equation}) becomes the following NLSE for
stationary states
\begin{equation}
-\frac{\alpha^2}{2}\frac{d^2 u^{1/2}}{d\xi^2
}~+~{\cal U}\left[u\right]u^{1/2}~=~E_0 u^{1/2}~~~,
\label{stationary-gnlse}
\end{equation}
where $E_0=c_0+u_{0}^{2}/2$. This equation admits
solitary solutions whose form is the squared modulus of
(\ref{envelope-soliton-quartic}). It follows that for any
\begin{equation}
Q_0~<~0~~~,
\label{Q-zero-condition}
\end{equation}
under condition (\ref{A-zero-condition}), solitary-wave solutions of
(\ref{mnlse-anti-cubic}) can be directly constructed from both
(\ref{Psi-solution}) and (\ref{envelope-soliton-quartic}) as
\begin{eqnarray}
& &\Psi_{{\pm}} (x,s)=\sqrt{\overline{u}~\left[1~+~\epsilon~\mbox{sech}
\left({\xi\over\Delta}\right)\right]}~{\times} \nonumber \\ & {\times} &
\exp\left\{\frac{i}{\alpha}
\left[\phi_0 -\left(E_0 + {u_{0}^{2}\over 2}\right)s+u_0 x
~{\pm}\sqrt{2|Q_0|}~
\int {d\xi \over\sqrt{\overline{u}~\left[1~+
~\epsilon~\mbox{sech}
\left(\xi /\Delta\right)\right]}}\right]\right\},
\label{Psi-solution-1}
\end{eqnarray}
where, in principle, according to Ref. \cite{2},
$\epsilon$ should be taken in the following range
\begin{equation}
-1~<\epsilon~\leq 1~~~,
\label{epsilon-condition}
\end{equation}
which excludes the standard "dark" solitary waves
($\epsilon
=-1$), namely the condition for which the modulus of $\Psi$
vanishes at $\xi = 0$. \newline Actually, the direct
substitution of $u=|\Psi|^2$ given by
(\ref{Psi-solution-1}) into the eigenvalue equation
(\ref{stationary-gnlse}) allows us to find
\begin{equation}
\epsilon~=~1~~~,
\label{epsilon-value}
\end{equation}
\begin{equation}
\overline{u}~=~-{3 q_1\over 8 q_2}~~~,
\label{u-overline}
\end{equation}
\begin{equation}
q_1~>~0~~~,~~~~~~q_2~<~0~~~,
\label{q1-q2-conditions}
\end{equation}
\begin{equation}
E_0~=~-{15 q_{1}^{2}\over 64 q_2}~~~,
\label{E-zero}
\end{equation}
\begin{equation}
\Delta=~{2|\alpha|\over q_1}\sqrt{2|q_2|\over 3}~~~.
\label{Delta}
\end{equation}
Consequently, solution (\ref{Psi-solution-1}) can be cast as
\begin{eqnarray}
&&\Psi_{{\pm}} (x,s)=\sqrt{{3 q_1\over 8 |q_2|}~\left[1~+~\mbox{sech}
\left({\xi\over\Delta}\right)\right]}~{\times} \nonumber \\ &{\times}&\exp\left\{\frac{i}{\alpha}
\left[\phi_0-\left({15 q_{1}^{2}\over 64 |q_2|} + {u_{0}^{2}\over 2}\right)s+u_0 x
~{\pm}\sqrt{2|Q_0|}~{16 |\alpha||q_2|\over 3q_{1}^{2}}\sqrt{2|q_2|\over
3}\left({\xi\over\Delta}-
\mbox{tanh}\left({\xi\over 2\Delta}\right)\right)\right]\right\}~~~,\nonumber \\
\label{Psi-solution-2}
\end{eqnarray}
where $u_0$ is a fully arbitrary soliton velocity.
According to the classification of the solitary waves
given in Ref. \cite{2}, the (\ref{Psi-solution-2})
represents an upper-shifted bright envelope solitonlike
solution of Eq. (\ref{mnlse-anti-cubic}), provided that
the coefficients $Q_0$, $q_1$ and $q_2$ satisfy the
conditions (\ref{Q-zero-condition}) and
(\ref{q1-q2-conditions}), respectively. It is clear that
Eq. (\ref{epsilon-value}), which does not contradict
condition (\ref{epsilon-condition}), implies that also
gray solitary solutions do not exist in the solution form
(\ref{Psi-solution-1}).\newline In conclusion, in this
paper, by using a recently developed method for solving a
wide family of MNLSE with high-order nonlinearities
\cite{1}-\cite{3} on the basis of the knowledge of the
solution of the associated MKdVE, an upper-shifted bright
envelope solitonlike solution of a MNLSE, containing,
besides the standard cubic and quintitic nonlinearities,
an "anti-cubic" nonlinear term (see Eq.
(\ref{mnlse-anti-cubic})~). Our analysis has shown that,
once the form of the envelope solution of
(\ref{mnlse-anti-cubic}) is taken according to
(\ref{Psi-solution-1}), dark and gray envelope solitary
solutions do not exist that, in contrast, exist in case
the "anti-cubic" term is missing.

\end{document}